\documentclass{ws-ijmpa}

\begin{document}

\markboth{J.~Gegelia and G.~Japaridze}
{On renormalizability of the effective field theory of massive Yang-Mills fields}

%
\catchline{}{}{}{}{}
%

\title{On renormalizability of the effective field theory of massive Yang-Mills fields}

\author{J.~Gegelia}

\address{Institut f\"ur Theoretische Physik II, Ruhr-Universit\" at Bochum\, Address\\
D-44780 Bochum, Germany
\\
Tbilisi State University, 0186 Tbilisi, Georgia
\\
Jambul.Gegelia@tp2.ruhr-uni-bochum.de}

\author{G.~Japaridze}

\address{Physics Department, Clark Atlanta University, 223 James P. Brawley Dr., SW\\
Atlanta, Georgia 30314, USA\\
gjaparidze@cau.edu}

\maketitle


\begin{abstract}
The effective field theory of massive Yang-Mills fields interacting with fermions is considered. Perturbative renormalizability in the framework of effective field theory is shown. It is argued that the limit of vanishing vector boson mass leads to massless gauge effective field theory. Possible relevance for the solution to the strong CP problem is discussed.

\end{abstract}

\ccode{PACS numbers: 11.10.Gf, 11.15.Tk}

\section{Introduction}

The modern point of view on the Standard Model  which is widely accepted as an established
theory of strong, electromagnetic and weak interactions treats the Standard Model as an effective field theory
(EFT)\cite{Weinberg:mt}. Traditional interpretation of renormalizability is replaced by the renormalizability in the sense of EFT, when the divergences are
absorbed in the redefinition of an infinite number of parameters of the effective Lagrangian.
The coupling constants of non-renormalizable interactions are suppressed by powers of some
large scale which makes the contributions of these interactions
negligible for lower energies leaving us with the predictive power of the
theory up to given accuracy. In other words, all the divergences should be removable from physical quantities by redefining {\it an infinite number} of parameters of the effective Lagrangian.
Hence the traditional renormalizability, which led to the gauge theories of the weak and strong interactions,
is no longer considered as a fundamental property. However,
it is very likely that for the self-interacting massive vector bosons, by demanding the
consistency with constraints and the perturbative renormalizability
in the sense of EFT, we are led to a gauge-invariant Lagrangian up to mass terms
(for a self-consistent EFT of self-interacting triplet of vector bosons see Ref.\cite{Djukanovic:2010tb}).

In this work we consider an EFT of massive $SU(N)$ Yang-Mills fields
interacting with fermions, i.e. we add a hard mass term to the most general $SU(N)$ gauge invariant Lagrangian.
To address the issue of the renormalizability, using the methods of
Refs\cite{'tHooft:1971fh,Slavnov:1972fg,Taylor:1971ff} we analyze the symmetries of the effective action and derive
the Slavnov-Taylor identities\cite{Slavnov:1972fg,Taylor:1971ff}. The symmetries of the effective action show the renormalizability of the considered EFT.
It turns out that in the framework of the perturbation theory the massless limit (mass of vector boson $\to 0$) does not exist. However, in a non-perturbative regime it could exist and the limit may reproduce the massless Yang-Mills theory\cite{Vainshtein:1971ip}.
Unlike the case of Ref.\cite{Vainshtein:1971ip}, the considered EFT is free of ultraviolet divergences and thus considering the limit mass of vector boson $\to 0$ in generating functional is a well-defined procedure and in the non-perturbative regime the massless limit exists. In case when the renormalized couplings with the negative mass dimensions are suppressed by inverse powers of some large scale, the considered EFT leads to the standard QCD\cite{Fritzsch:1973pi} in high-energy regime, i.e. for energies which are much larger than the  vector meson mass but still much smaller than the large scale.

The paper is organized as follows: in section 2 we introduce the model and consider the symmetries of the effective action. The analysis shows that the effective action has all the symmetries of the effective Lagrangian and  the vector boson mass does not get loop corrections. This guarantees the renormalizability of the considered effective field theory. In section 3 we demonstrate the renormalizability in the example of the renormalization of the coupling constant of the leading order Lagrangian. In section 4 we consider the vanishing mass limit and show that while in perturbation theory the limit does not exist, in non-perturbative regime the limit is well-defined and coincides with the massless EFT. In section 5, we discuss the results and mention an interesting possibility that the considered effective field theory may be free from the $U(1)$ and the strong CP problem and simultaneously reproduces QCD in high-energy regime.

\section{Symmetries of the effective action of the model}
We consider an EFT Lagrangian of the massive Yang-Mills vector fields
interacting with fermions given by
\begin{align}
{\mathcal L} &= {\mathcal L}_{0}\,+\,{\mathcal L}_{1}\,=\,-\frac{1}{4}\,\sum_a\, G^{a \mu\nu}
G^a_{\mu\nu}+\frac{M^2}{2}\,\sum_a B^\mu_a B_{a \mu} +
\theta\,\epsilon^{\mu\nu\alpha\beta} G^a_{\mu\nu}
G^a_{\alpha\beta}\nonumber\\
& + \sum_{q,j} \bar \psi_q^j \left( i
\partial \hspace{-.6em}/\hspace{.1em}-m_q\right)
\psi_q^j+g\sum_{q,i,k,a} \bar \psi_q^i \gamma_{\mu} t^{a}_{ik}\psi_q^{k} B_{a}^{\mu}+
{\mathcal L}_{1}\,,
 \label{lagrangianQCD01}
\end{align}
where $G^a_{\mu\nu}=\partial_\mu B^a_\nu-\partial_\nu B^a_\mu + g
f^{abc} B_\mu^bB_\nu^c$ and $t^a$ and $f^{abc}$, respectively, are the generators and the totally
antisymmetric structure constants of the $SU(N)$ group.
The summation over $q$ covers different flavors of fermions. All possible Lorentz and gauge invariant terms (an infinite number of them) with
coupling constants with negative mass-dimensions are contained in ${\mathcal L}_1$.
It is assumed that renormalized
coupling constants with negative mass dimensions
are suppressed by powers of some large scale.

To study renormalizability of the system described by effective Lagrangian (\ref{lagrangianQCD01}) we introduce Green's functions
\begin{equation}
\langle T B_\mu^a(x_1)\cdots \bar\psi(x_i)\cdots \psi(x_n)\rangle=\frac{i^{-n}\,\delta^n
Z}{\delta J^a_\mu(x_1)\cdots \delta\xi(x_i)\cdots \delta\bar\xi
(x_n)}|_{J=\xi =\bar\xi =0}\,. \label{GRF}
\end{equation}
where the generating functional of Green's functions for considered EFT model has the
form (for the Lagrangian (\ref{lagrangianQCD01}) we have derived the generating functional rigorously using the standard canonical quantization in Hamilton formalism with constraints present\cite{gitman})
\begin{equation}
Z[J^a_\mu,\xi,\bar\xi] = \int {\mathcal D} B\,{\mathcal D} \psi\,{\mathcal D}
\bar\psi\,\exp \left\{ i \int d^4 x \,\left[ {\mathcal
L}(x)+J^{a\mu}B^a_\mu +\xi \bar\psi + \bar \xi\psi
\right]\right\}, \label{GFFinal0}
\end{equation}
In (\ref{GFFinal0}) the flavor and group indices are not shown for fermion fields. As the Lagrangian (\ref{lagrangianQCD01}) does not mix the fermions of different flavors, the corresponding index is suppressed throughout.

Following Refs.\cite{'tHooft:1971fh,Slavnov:1972fg,Taylor:1971ff} let us make the
change of integration variables in Eq.~(\ref{GFFinal0}) by performing an
infinitesimal gauge transformation
\begin{align}
B^a_\mu(x) & \to  B^a_\mu(x) + g\,f^{abc} B_\mu^b(x)
\phi^c(x)+\partial_\mu \phi^a(x)\,,\nonumber\\
\bar\psi(x) &\to  \bar\psi(x)-i\,g\,\bar\psi\, t^a \phi^a(x) \,,\nonumber\\
\psi(x) &\to \psi(x)+i\,g\,t^a
\psi\,\phi^a(x)\,,\label{gaugetransf}
\end{align}
where $\phi^a(x)$ are arbitrary infinitesimal functions. Generating functional $Z[J^a_\mu,\xi,\bar\xi]$ is invariant under transformations (\ref{gaugetransf}), therefore its variation vanishes. The measure and the Lagrangian ${\mathcal L}$, except the vector boson mass term, are invariant; the mass term of the vector boson and the source terms transform non-trivially. Thus, invariance of the generating functional under transformations of Eq.~(\ref{gaugetransf}) results in the identity
\begin{align}
&\int d^4 z \biggl[ M^2 \langle B^{c \mu}(z)\rangle_{S}\partial_\mu \phi^c(z)-\frac{\delta\Gamma[B,\bar\psi,\psi]}{\delta B^c_\mu(z)} \partial_\mu\phi^c(z) \nonumber\\
&- g\,f^{abc} \frac{\delta\Gamma[B,\bar\psi,\psi]}{\delta B^a_\mu(z)}\,\langle B^b_\mu(z)\rangle_S\phi^c(z)
+ i\,g\,\frac{\delta\Gamma[B,\bar\psi,\psi]}{\delta \bar\psi(z)}\langle \bar\psi(z)\rangle_S\,t^c \phi^c(z)\nonumber\\
&- i\,g\,\frac{\delta\Gamma[B,\bar\psi,\psi]}{\delta\psi(z)} \,t^c\langle \psi(z)\rangle_S \phi^c(z) \biggr]\,=\,0
\label{IdentityRewritten}
\end{align}	
where
\begin{align}
Z[J,\xi,\bar\xi] \langle O(z)\rangle_S &\equiv \int {\mathcal D} B\,{\mathcal D} \psi\,{\mathcal D} \bar\psi\,O(z)\exp\left\{i \int d^4 x \,
\left[{\mathcal L}(x)+J^{a\mu}B^a_\mu +\xi \bar\psi + \bar\xi\psi \right]\right\}
\label{<O>definition}
\end{align}
and $\Gamma[B,\bar\psi,\psi]$ is the effective action, defined in the usual way as a Legendre transform of the generating functional\cite{Weinberg:mt}. According to Eq.~(\ref{IdentityRewritten}), the variation of the effective action under gauge
transformations is equal to the variation of the mass term, i.e. effective action consists of a gauge invariant part and the non invariant mass term of the vector bosons. It follows from the first term of Eq.~(\ref{IdentityRewritten}) that the vector boson mass does not get quantum corrections. If we drop the mass term, the rest of the quantum effective action is gauge invariant, exactly as the effective Lagrangian is.

The renormalization procedure in the framework of effective field theory presents effective
Lagrangian as a Taylor series expansion in derivatives of fields; divergences are absorbed order-by-order in this expansion\cite{Weinberg:mt}. In momentum space this means that the Green's functions are expanded in powers of momenta and the divergences are absorbed in fields and parameters of the effective Lagrangian. When the symmetry of the quantum effective action is the same as the symmetry of the classical action defined by effective Lagrangian, renormalization of the fields and (an infinite number of) parameters of the effective Lagrangian removes all divergences in loop expansion\cite{Weinberg:mt}

 Divergences of all loop diagrams appearing in any local quantum field theory
can be subtracted in a self-consistent way using Zimmerman's forest formula, and these subtractions  can be realized by including corresponding local counter terms in the Lagrangian (see e.g. Refs.\cite{Weinberg:mt},\cite{Collins:1984xc}). Due to the symmetries of the effective action all the counter terms in our effective field theoretical
model satisfy the constraints of gauge invariance. As the effective Lagrangian contains {\it all} the terms which are invariant under gauge transformations, all these counter terms can be absorbed by redefining corresponding parameters and fields of the effective Lagrangian\cite{Weinberg:mt}. In other words, the effective quantum field theory defined by Lagrangian (\ref{lagrangianQCD01}) is renormalizable. The vector boson mass term does not get quantum corrections and therefore the mass parameter will be renormalized only due to the renormalization of the vector field, i.e. $M^2 B_\mu^a B^{a \mu}=M^2 Z_B B_{R\,\mu}^{a} B_R^{a \mu} = M^2_R B_{R\,\mu}^{a} B_R^{a \mu}$, where $B_R^{a\mu}$ and $M_R$ are the renormalized field and the mass and $Z_B$ is the renormalization factor of the vector field.

\section{Slavnov-Taylor identities and renormalization}
Here we demonstrate renormalizability analyzing Slavnov-Taylor identities.
Equating the coefficient of
$\phi^a(z)$ in Eq.~(\ref{IdentityRewritten}) to zero we obtain
\begin{align}
& \int {\mathcal D} B\,{\mathcal D} \psi\,{\mathcal D} \bar\psi\,\exp \left\{i
\int d^4 x \,\left[ {\mathcal L}(x)+J^{a\mu}B^a_\mu +\xi \bar\psi + \bar
\xi\psi \right]\right\}\nonumber\\
& \ \ \ \times  \biggl[ M^2 \partial_\mu B^{c \mu}(z)+\partial_\mu
J^{c \mu}(z)- g\,f^{abc} J^{a \mu}(z)\,B^b_\mu(z)\nonumber\\
& \ \ \ +i\,g\,\xi(z)\,\bar\psi(z)\,t^c - i\,g\,\bar \xi(z)\,t^c
\psi(z) \biggr]=0\,. \label{STI}
\end{align}
	Equation (\ref{STI}) generates Slavnov-Taylor identities. Below we omit the lengthy calculations and just list the results for some particular cases necessary to discuss  renormalizability on the example of $BBB$, $BBBB$ and $B\bar \psi \psi$ vertex functions.

The first identity is obtained by differentiating Eq.~(\ref{STI}) with respect to $J^a_\alpha(x)$ and setting $J^b_\nu=\xi=\bar\xi=0$. In momentum space it reads
\begin{equation}
M^2 p^\mu \,i\,\delta^{ac} S_{\mu\nu}(p) =
i\,\delta^{ac}\,p_{\nu}\,,\label{STIPropM}
\end{equation}
where $i\,\delta^{ac} S_{\mu\nu}(p)$ is the dressed propagator of
the vector boson. It can be written as
\begin{equation}
i\,\delta^{ab} S_{\mu\nu}(p) = -i\, \delta^{ab}\,\frac{g_{\mu\nu}-
p_\mu p_\nu \left[1+ \Pi(p^2)\right]/M^2}{p^2 \left[1+\Pi(p^2)\right]-M^2+i\,0^+}\,.\label{dressedprop}
\end{equation}
Here $\Pi^{ab}_{\mu\nu}(p)$, the sum of all one-particle-irreducible diagrams contributing in the vector boson two-point function is presented as
\begin{equation}
i\,\Pi^{ab}_{\mu\nu}(p)=i\,\delta^{ab}\left[ -g_{\mu\nu}\,p^2+p_\mu
p_\nu\right]\Pi(p^2)\,, \label{VSEparNew}
\end{equation}
and we have verified that the tensor structure of $\Pi^{ab}_{\mu\nu}(p)$ given by (\ref{VSEparNew}) indeed follows from (\ref{STIPropM}). Expanding $\Pi(p^2)$ in powers of $p^2$, the renormalization of the vector
field which has to absorb the corresponding divergence in the dressed propagator can be defined as:
\begin{equation}
Z_B = \frac{1}{1+\Pi(0)}\,.
\label{ZB}
\end{equation}

Another identity is obtained by differentiating Eq.~(\ref{STI}) with respect to $\bar\xi$ and $\xi$
and setting $J^b_\nu=\xi=\bar\xi=0$. In momentum space we have:
\begin{equation}
M^2\, (p_f-p_i)_\mu\,G^{c \mu}_{ji}(p_f,p_i) = g
S_{ki}(p_i)\,t^c_{kj}-g S_{jk}(p_f)\,t^c_{ik}\,, \label{SLTIVFF}
\end{equation}
where $G^{c \mu}_{ji}(p_f,p_i)$ is the $B\bar\psi\psi$ Green's function and
\begin{equation}
iS_{ij}(p) =\frac{i\,\delta_{ij}}{p\hspace{-.45em}/\hspace{.1em}-m-\Sigma (p\hspace{-.45em}/\hspace{.1em})}
\label{gia}
\end{equation}
is the (dressed) fermion propagator with $-i\,\delta_{ij}\Sigma(p\hspace{-.45em}/\hspace{.1em})$ being the sum of all one-particle irreducible diagrams contributing in the fermion two-point function.

\noindent Renormalization constant of the fermion field which has to cancel the
corresponding divergence in the dressed propagator can be defined as follows:
\begin{equation}
Z_\psi = \frac{1}{1-\Sigma'(0)}\,,\;\;\;\Sigma'(0)\equiv \frac{\partial\Sigma(x)}{\partial x}|_{x=0}\,.
\label{ZPsi}
\end{equation}
By writing
\begin{equation}
G^{c \mu}_{ji}(p_f,p_i) =
i\,S^{\mu\alpha}(p_f-p_i)\,i\,S_{jj_1}(p_f)\,\Gamma^{c}_{j_1i_1
\alpha}(p_f,p_i)\,i\,S_{i_1i}(p_i), \label{VFFvertexdef}
\end{equation}
it follows from Eq.~(\ref{SLTIVFF}):
\begin{equation}
(p_f-p_i)_\mu\,\Gamma_{ji}^{c \mu}(p_f,p_i) = i\,g\,
t^c_{ji} \left\{\left[ p_f\hspace{-.87em}/\hspace{.1em} -m -\Sigma(p_f\hspace{-.87em}/\hspace{.1em})\right]
-\left[ p_i\hspace{-.72em}/\hspace{.1em} -m -\Sigma(p_i\hspace{-.72em}/\hspace{.1em})\right]\right\}\,.
\label{SLTIVFFVertexFunctio}
\end{equation}
	Expanding the vertex function in powers of momenta
\begin{equation}
\Gamma_{ji}^{c \mu}(p_f,p_i) = i\,g\,
t^c_{ji} \gamma^\mu [1+\Gamma(0)]+{\mathcal O}(p)\,,
\label{vertexpar}
\end{equation}
substituting in Eq.~(\ref{SLTIVFFVertexFunctio}) and comparing the coefficients of leading
orders in momenta we obtain
\begin{equation}
\Gamma(0) = -\Sigma'(0)\,.
\label{id1}
\end{equation}
	By multiplying Eq.~(\ref{vertexpar}) with $Z_\psi Z_B^{1/2}$ and taking into account
Eqs.~(\ref{ZPsi}) and (\ref{id1}) we obtain for the renormalized vertex function
\begin{equation}
\Gamma_{ji}^{c \mu, R}(p_f,p_i) = i\,g\,Z_\psi Z_B^{1/2}\,
t^c_{ji} \gamma^\mu [1+\Gamma(0)]+{\mathcal O}(p)=i\,g\,Z_B^{1/2}\,
t^c_{ji}\gamma^\mu +{\mathcal O}(p)\,,
\label{vertexparRen}
\end{equation}
which defines the renormalized coupling constant
\begin{equation}
g^R = g\,Z_B^{1/2}\,.
\label{gR1}
\end{equation}

The next identity is obtained by differentiating Eq.~(\ref{STI}) with respect to $J^a_\alpha(x_1)$
and $J^b_\beta(x_2)$ and setting $J^n_\nu=\xi=\bar\xi=0$ which gives in momentum space
\begin{equation}
M^2\, p^\mu\,G_{\mu\nu\lambda}^{abc}(p,q,k) = i\,g\,f^{abc}
\left[ S_{\lambda\nu}(p+k)-S_{\lambda\nu}(p+q)\right]\,,
\label{SLTIVVV}
\end{equation}
where $G_{\mu\nu\lambda}^{abc}(p,q,k)$ is the Green's function of three vector bosons.
By writing
\begin{equation}
G_{\mu\nu\lambda}^{abc}(p,q,k) =
i\,S_{\mu\alpha}(p)\,i\,S_{\nu\beta}(q)\,i\,S_{\lambda\gamma}(k)\,\Gamma_{\alpha\beta\gamma}^{abc}(p,q,k),
\label{vertexdef}
\end{equation}
Eq.~(\ref{SLTIVVV}) reduces to
\begin{equation}
p^\mu\,\Gamma_{\mu\nu\lambda}^{abc}(p,q,k) = g\,f^{abc} \left[
S^{-1}_{\nu\lambda}(k)-S^{-1}_{\nu\lambda}(q)\right]\,, \label{SLTIVVVreduced}
\end{equation}
where the inverse to the propagator $S^{\mu\nu}(p)$ is given by
\begin{equation}
S^{-1}_{\mu\nu}(p) = -M^2 g_{\mu\nu}+\left(p^2 g_{\mu\nu}-p_\mu p_\nu\right)[1+\Pi(p^2)].
\label{vprinv}
\end{equation}

\noindent Up to terms quadratic in momenta we obtain from Eq.~(\ref{SLTIVVVreduced}) (using $k+p+q=0$)
\begin{equation}
\Gamma_{\mu\nu\lambda}^{abc}(p,q,-p-q) = g\,f^{abc} \left[
g^{\lambda\nu}(p^\mu+2\,q^\mu)-g^{\mu\nu} q^\lambda-g^{\lambda\mu} (p+q)^\nu
\right][1+\Pi(0)]
. \label{SLTIVVVVertexFunctio}
\end{equation}
\noindent
	By multiplying Eq.~(\ref{SLTIVVVVertexFunctio}) with $Z_B^{3/2}$ and taking into
account Eq.~(\ref{ZB}),\\ $Z_{B}\left[1+\Pi(0)\right]=1$, we obtain for the renormalized Green's function (up to terms quadratic in momenta)
\begin{equation}
\Gamma_{\mu\nu\lambda}^{abc,R}(p,q,-p-q)=g\,Z_B^{1/2}f^{abc} \left[
g^{\lambda\nu}(p^\mu+2\,q^\mu)-g^{\mu\nu} q^\lambda-g^{\lambda\mu} (p+q)^\nu
\right], \label{SLTIVVVVertexFunctioRen}
\end{equation}
which defines exactly the same renormalized coupling, $g^R = g\,Z_B^{1/2}$, as given in Eq.~(\ref{gR1}).

One more identity (for a four-point function) can be obtained by differentiating Eq.~(\ref{STI}) with respect
to $J^a_\alpha(x_1)$, $J^b_\beta(x_2)$ and $J^d_\gamma(x_3)$ and setting
$J^n_\nu=\xi=\bar\xi=0$.  By writing the
connected part of the Green's function of four vector bosons in momentum space as
\begin{equation}
\tilde G_{\alpha\beta\gamma\mu}^{abdc}(p,q,k,r) =
i\,S_{\alpha\alpha_1}(p)\,i\,S_{\beta\beta_1}(q)\,i\,S_{\gamma\gamma_1}(k)
\,i\,S_{\mu\mu_1}(r)\,\tilde\Gamma_{\alpha_1\beta_1\gamma_1\mu_1
}^{abdc}(p,q,k,r), \label{4vertexdef}
\end{equation}
identity for $\tilde\Gamma$ turns out to be
\begin{align}
r^\mu\,\tilde\Gamma_{\alpha\beta\gamma\mu}^{abdc}(p,q,k,r) & =
i\,g
f^{acm}\,S^{-1}_{\alpha\alpha_1}(p)\,S^{\alpha_1\alpha_2}(p+r)\,
\Gamma_{\alpha_2\beta\gamma}^{mbd}(p+r,q,k)\nonumber\\
& +  i\, g f^{bcm}\,S^{-1}_{\beta\beta_1}(q)\,S^{\beta_1\beta_2}(q+r)\, \Gamma_{\alpha\beta_2\gamma}^{amd}(p,q+r,k)\nonumber\\
& +  i\,g f^{dcm}\,S^{-1}_{\gamma\gamma_1}(k)\,S^{\gamma_1\gamma_2}(k+r)\,
\Gamma_{\alpha\beta\gamma_2}^{abm}(p,q,k+r)\,.
\label{SLTIVVVVVertexFunctio}
\end{align}
Up to the terms quadratic in momenta Eq.~(\ref{SLTIVVVVVertexFunctio})
leads to
\begin{align}
r^\mu\,\tilde\Gamma_{\alpha\beta\gamma\mu}^{abdc}(p,q,k,r) & =
-i\,g^2\Bigl\{ f^{acm}\,
f^{mbd}\left[g^{\alpha\beta}r^\gamma-g^{\alpha\gamma}r^\beta\right]
+  f^{bcm}\,f^{amd}\left[g^{\beta\gamma}r^\alpha-g^{\alpha\beta}r^\gamma\right]\nonumber\\ &+
f^{dcm} \,
f^{abm}\left[g^{\alpha\gamma}r^\beta-g^{\beta\gamma}r^\alpha\right]\Bigr\}[1+\Pi(0)]\,.
\label{SLTIVVVVVertexFunctio3}
\end{align}
By multiplying Eq.~(\ref{SLTIVVVVVertexFunctio3}) with $Z_B^{2}$ and taking into
account Eq.~(\ref{ZB}), $Z_{B}\left[1+\Pi(0)\right]=1$, we obtain for the renormalized Green's function (up to terms quadratic in momenta)
\begin{align}
r^\mu\,\tilde\Gamma_{\alpha\beta\gamma\mu}^{abdc,R}(p,q,k,r) & =
-i\,g^2 Z_B \Bigl\{ f^{acm}\,
f^{mbd}\left[g^{\alpha\beta}r^\gamma-g^{\alpha\gamma}r^\beta\right]\nonumber\\
&+  f^{bcm}\,f^{amd}\left[g^{\beta\gamma}r^\alpha-g^{\alpha\beta}r^\gamma\right] \nonumber\\
& +
f^{dcm} \,
f^{abm}\left[g^{\alpha\gamma}r^\beta-g^{\beta\gamma}r^\alpha\right]\Bigr\}\,,
\label{SLTIVVVVVertexFunctioR}
\end{align}
which again defines the same renormalized coupling as the one given in Eq.~(\ref{gR1}).

Hence,  in $BBB$, $BBBB$ and $B\bar \psi \psi$ vertex functions the same renormalized coupling constant, $g^R = g\,Z_B^{1/2}$ emerges. This result, together with the fact that the vector boson mass is renormalized only due to the renormalization of the vector field, as discussed in the end of the previous section, leads to the conclusion that the renormalization of the fields and parameters of the leading order Lagrangian ${\mathcal L}_{0}$ (explicitly shown in Eq.~(\ref{lagrangianQCD01})) absorbs all the corresponding divergences to all orders in loop expansion. Note that this applies to all loop diagrams, including those ones which are generated by interaction terms of ${\mathcal L}_1$. This demonstrates the general result obtained from the analysis of the effective action that our effective field theoretical model is renormalizable.

We have verified  the above identities and renormalizability in explicit calculations at one-loop order using  dimensional regularization.
\section{Vanishing mass limit}

In this section, in analogy with Ref.\cite{Vainshtein:1971ip},
we consider the limit $M\rightarrow0$ in
both perturbative and non-perturbative regimes.
Let us examine the Green's functions
\begin{equation}
G_N(x_1,\cdots,x_N) = \int {\mathcal D} B\,{\mathcal D} \psi\,{\mathcal D}
\bar\psi\,{\mathcal O}_1(x_1)\cdots {\mathcal O}_N(x_N)\,\exp \left\{i\int d^4 x \,{\mathcal
L}(x)\right\}\,, \label{GFFinal}
\end{equation}
where ${\mathcal O}_i$ are invariant under the $SU(N)$ gauge transformations (\ref{gaugetransf}). Following Ref.\cite{Slavnov:1970tk}, we use the Faddeev-Popov trick and insert in the righthand side of Eq.~(\ref{GFFinal}) the identity
\begin{equation}
1=\Delta_F[B_\mu^a]\int {\mathcal D}U\,\delta[\partial^\mu B^{a U}_\mu],\quad t^a B^{a U}_\mu = U t^a B^a_\mu U^{-1}-\frac{i}{g}\,\partial_\mu U \,U^\dagger
\label{FPI}
\end{equation}
where  $U$ is an element of the $SU(N)$ group. Next we perform the gauge transformation in the path integral which transforms $B^{U}_\mu$ into $B_\mu$.  The only term which is not invariant under this transformation is the vector boson mass term
${\rm Tr}\left\{ B_\mu B^{\mu}\right\}$ which transforms as follows
\begin{equation}
\frac{M^2}{4}\,{\rm Tr}\left\{ B^{U}_\mu B^{\mu U}\right\} = \frac{1}{4}\,{\rm Tr}\left\{\left( MB^\mu+\Sigma^\mu\right)\left(MB_\mu+\Sigma_\mu\right)\right\},
\label{massttransformed}
\end{equation}
where
\begin{equation}
\Sigma_\mu =\frac{iM}{g}\,\partial_\mu U U^{-1}.
\label{SigmaDef}
\end{equation}
For the Green's functions we obtain
\begin{align}
G_N(x_1,\cdots,x_N) & =  \int {\mathcal D}U\,{\mathcal D} B\,{\mathcal D} \psi\,{\mathcal D}
\bar\psi\,\Delta_F[B_\mu^a]\,\delta[\partial^\mu B^{a}_\mu]\nonumber\\
& \times  {\mathcal O}_1(x_1)\cdots {\mathcal O}_N(x_N)\,\exp \left\{i\int d^4 x \,{\mathcal
L}_{\rm \Sigma}\right\}, \label{GFFinalTransformed}
\end{align}
where
\begin{eqnarray}
{\mathcal L}_{\Sigma} & = &{\mathcal L}
+\frac{M}{2}\,{\rm Tr}\left\{\Sigma^\mu B_\mu\right\}+\frac{1}{4}\,{\rm Tr}\left\{\Sigma^\mu \Sigma_\mu\right\}.
\label{BUdef}
\end{eqnarray}
We parameterize $U$ in terms of new fields $\sigma(x)$ and $y^{a}(g,\,M,\,x)$ as follows
\begin{align}
U & =  \frac{W}{({\rm det}\,W)^{1/N}}\,,\nonumber\\
W & =  \sqrt{1-\frac{1}{2N} g^4\,y(g,M)^2-\frac{32 g^2
   M^2 \sigma^2}{N\left(g^2 \sigma^2+4 M^2\right)^2}}+g\,t^a
   \left[g\,y^a(g,M)+\frac{8\,i M \sigma^a}{g^2 \sigma^2+4
   M^2}\right],
\label{Upar}
\end{align}
where we have suppressed the argument $x$ in functions $y^a$ and $\sigma^a$. By demanding that $W$ is unitary we obtain that $U$ is unitary and unimodular.

From the unitarity condition for $W$ we obtain the following equation for $y^a(g, M)$
\begin{align}
0 & =
\frac{8 M \left[8 M d^{abc}\sigma^a \sigma^b+g \left(g^2 \sigma^2
+4 M^2\right) f^{abc} y^a(g,M)\sigma^b \right]}{\left(g^2 \sigma^2+4 M^2\right)^2}\nonumber\\
& +  g^2 d^{abc} y^a(g,M) y^b(g,M)+2
   y^c(g,M) \sqrt{4-\frac{2 g^4 y(g,M)^2}{N}-\frac{128
    g^2 M^2 \sigma^2}{N\left(g^2 \sigma^2+4 M^2\right)^2}}\,,
\label{Uequation}
\end{align}
where $f^{abc}$($d^{abc}$) are totally antisymmetric(symmetric) structure functions of the $SU(N)$ group and  $\sigma^2\equiv\sum_{a=1}^N \sigma^a \sigma ^a$,  $y^2\equiv\sum_{a=1}^N y^ay ^a$.

Eq.~(\ref{Uequation}) can be solved in the framework of  perturbation theory 
 as an expansion in powers of $g$. When substituting that solution in the effective Lagrangian it follows (in agreement with the well known result, see e.g. Ref.\cite{Vainshtein:1971ip}) that the limit $M\to0$ does not exist in the perturbation theory.

On the other hand Eq.~(\ref{Uequation}) can be solved perturbatively for $y^a$ as a Taylor series in mass of vector boson, $M$:
\begin{equation}
y^a(g,M) = -\frac{16 M^2 d^{abc}\sigma^b \sigma^c}{g^4 \sigma^4}+{\mathcal O}(M^3).
\label{eqM0}
\end{equation}
This results in the following expression for the group element $U$:
\begin{align}
U(x) & =  1+ {\mathcal O}(M),\nonumber\\
\partial_\mu U(x) & =  \frac{8\,i\,M\,t^a \left(\sigma^2 \partial_\mu\sigma^a-2\, \partial_\mu\sigma^b\sigma^b\sigma^a\right)}{g \,\sigma^4}+{\mathcal O}(M^2).
\label{derinM0}
\end{align}
From Eqs.~(\ref{SigmaDef}), (\ref{eqM0}), and (\ref{derinM0}), it follows that $\Sigma_\mu$ vanishes in $M\to 0$ limit and hence $\lim_{M\rightarrow 0}{\mathcal L}_{\rm \Sigma}={\mathcal L}(M=0)$ where ${\mathcal L}$ is given in Eq.~(\ref{lagrangianQCD01}).  In other words, the limit $M\to0$ does exist in our EFT in a non-perturbative regime. As the remaining integration over the $U$ field can be absorbed in the normalization it follows that the non-perturbative
Green's functions of Eq.~(\ref{GFFinalTransformed}) converge to the ones of the massless theory. Note that the effective Lagrangian (\ref{lagrangianQCD01}) contains an infinite number of interaction terms so that all ultraviolet divergences can be absorbed in the redefinition of the parameters which means that the limit $M\to 0$ is well-defined in the framework of the presented EFT.
\section{Summary and discussion}
In this work we considered an effective field theory of massive Yang-Mills fields interacting with massive fermions.
An effective Lagrangian consists of all local terms which are compatible with the Lorentz invariance and $SU(N)$ gauge symmetry, plus a globally invariant mass term of the vector bosons.
To address the issue of renormalizability of the considered EFT, we investigated the symmetries of the effective
action. It turns out that up to the vector boson mass term the quantum effective action has exactly the same $SU(N)$ gauge symmetry as the action generated by the effective Lagrangian (\ref{lagrangianQCD01}), and the vector boson mass does not get loop corrections. This accounts to the renormalizability requirement for effective field theory\cite{Weinberg:mt}, i.e.  considered effective field theory set up by Lagrangian (\ref{lagrangianQCD01}) is renormalizable.
We have also derived the Slavnov-Taylor identities and demonstrated the renormalizability of the model by explicit calculations in one-loop order.

	If the effective Lagrangian is treated perturbatively by expanding in powers of the coupling
constants, then the limit of vanishing vector boson mass $M\to 0$ does not exist. On the other hand, in this limit the non-perturbative Green's functions converge to the ones of the massless theory. Unlike the massive Yang-Mills theory considered in Ref.\cite{Vainshtein:1971ip}, our EFT model is free of ultraviolet divergences; hence our result is self-consistent.
	As a consequence of this non-perturbative existence of the massless limit, for energies
much larger than the vector boson mass, effective theory can be well approximated by the massless
effective theory. In case the renormalized coupling constants with negative mass dimensions are  suppressed by
some scale, which is much bigger than the considered energies, we arrive to the standard QCD (for $SU(3)$ case). In the light
fermion sector at low energies, where the massive bosons can be
integrated out, the leading-order effective Lagrangian
describes (a generalized version of) the model of Nambu and
Jona-Lasinio\cite{Nambu:1961tp}.

    Finally, let us mention an intriguing scenario leading to a different viewpoint on a strong CP problem. The mass term of the vector bosons excludes the configurations with
slowly decaying (instanton-like) boundary conditions. Hence, as the CP-violating $\theta$-term $\theta\,\epsilon^{\mu\nu\alpha\beta} G^a_{\mu\nu}
G^a_{\alpha\beta}$ can be written as a total derivative, it will not have any effect on physical quantities. This solves the strong CP problem
without introducing additional scalar field(s)\cite{Peccei:1977hh,Peccei:1977ur}.
    On the other hand, the $U(1)$ problem, which has been resolved by  instantons\cite{'tHooft:1986nc}, re-appears in the effective theory. However, due to the stronger UV divergences, higher loop diagrams could give non-trivial contributions in the anomaly of the singlet axial current so that the divergence of this current can no longer be written as a total derivative. In other words, in the framework of the considered EFT model  the $U(1)$ problem may be solved and at the same time a strong CP violating term will factor out from the dynamics. This speculative scenario requires a more detailed investigation which is beyond the scope of the present paper.

\section*{Acknowledgment}
The authors would like to thank D.~Djukanovic for discussions and comments on the manuscript.
J.G.~acknowledges the support of
the Deutsche Forschungsgemeinschaft and Georgian National
Foundation grant GNSF/ST08/4-400.


\end{document}